\documentclass[a4paper]{llncs}

\usepackage{graphicx}
\usepackage{multirow}
\usepackage{url}
\usepackage{color}
\usepackage{rotating}
\usepackage{listings}
\usepackage{microtype}

\begin{document}

\title{Space-Time Diagram Generation for Profiling Multi Agent Systems}

\author{Dinh Doan Van Bien \and David Lillis \and Rem W. Collier}

\institute{School of Computer Science and Informatics \\ University College Dublin \\
\email{dinh@doanvanbien.com, \{david.lillis,rem.collier\}@ucd.ie}}

\maketitle

\begin{abstract}
Advances in Agent Oriented Software Engineering have focused on the provision of frameworks and toolkits to aid in the creation of Multi Agent Systems (MASs). However, despite the need to address the inherent complexity of such systems, little progress has been made in the development of tools to allow for the debugging and understanding of their inner workings.

This paper introduces a novel performance analysis system, named AgentSpotter, which facilitates such analysis. AgentSpotter was developed by mapping conventional profiling concepts to the domain of MASs. We outline its integration into the Agent Factory multi agent framework.
\end{abstract}

\section{Introduction} \label{sect:intro}

Recent developments in the area of Multi Agent Systems (MASs) have been concerned with bridging the gap between theory and practice, by allowing concrete implementations of theoretical foundations to be built and deployed. However, the dearth of agent-specific development and debugging tools remains a significant obstacle to MASs being adopted in industry on a large scale.

While some simple debugging and logging tools exist for MAS analysis, these tend not to aid in reasoning about large-scale system when viewed at the high agent-oriented abstraction level. Such tools typically allow for traditional debugging actions such as state stepping and breakpoint insertion.

One popular performance analysis technique is known as \emph{profiling}. Profiling is based on the observation that the majority of the execution time of a program can be attributed to a small number of \emph{bottlenecks} (or \emph{hot spots}). By improving the efficiency of these portions of a program, overall performance can be dramatically improved. Profiling was initially introduced by Donald E. Knuth in an empirical study conducted on FORTRAN programs \cite{Knuth1971}. Since then, the technique has been successfully applied to a variety of languages, platforms and architectures.

The aim of this paper is to apply the principles of traditional profiling systems in a multi agent environment, so as to facilitate the developers of MASs in debugging their applications by gaining a better understanding of where the bottlenecks exist and performance penalties are incurred.

This paper is organised as follows: Section~\ref{sect:related} provides a brief overview of existing tools aimed at aiding in the analysis of MASs. In Section~\ref{sect:overview}, we introduce the AgentSpotter profiling system, with particular focus on outlining a conceptual model for generic MAS profiling. A concrete implementation of this work, aimed at the Agent Factory MAS framework, is outlined in Section~\ref{sect:integration}. Section~\ref{sect:spacetime} presents the space-time diagram produced by AgentSpotter in more detail, with an evaluation of its usefulness given in Section~\ref{sect:evaluation}. Finally, Section~\ref{sect:conclusions} presents our conclusions and ideas for future work.

\section{Related Work} \label{sect:related}

In designing a profiler for MASs, the features that tend to be present in traditional profilers for non-MAS applications must be identified. It is also necessary to examine those debugging and analysis tools that already exist for MASs.

The motivation behind the use of profiling on computer applications is clearly outlined in Knuth's observation that ``less than 4\% of a program accounts for more than half of its running time'' \cite{Knuth1971}. This statement implies that a developer can achieve substantial increases in performance by identifying and improving those parts of the program that account for the majority of the execution time. The key aim of profilers is to identify these bottlenecks.

Another observation leading to the widespread adoption of profilers as debugging tools is that there frequently exists a mismatch between the actual run-time behaviour of a system and the programmers' mental map of what they expect this behaviour to be. Profilers are useful in enlightening developers to particular aspects of their programs that they may not otherwise have considered.

A traditional profiler typically consists of two logical parts. Firstly, an \emph{instrumentation apparatus} is directly weaved into the program under study or run side-by-side to gather and record execution data. Secondly, a \emph{post-processing system} uses this  data to generate meaningful performance analysis listings or visualisations.

In the traditional software engineering community, historical profilers such as gprof~\cite{Graham1982} or performance analysis APIs like ATOM~\cite{Srivastava1994} and the Java Virtual Machine Tool Interface (JVMTI) \cite{jvmti2004} have made performance analysis more accessible for researchers and software engineers. However, the MAS community does not yet have general access to these types of tools.

Unique amongst all of the mainstream MAS development platforms, Cougaar is alone in integrating a performance measurement infrastructure directly into the system architecture \cite{Helsinger2004}. Although this is not applicable to other platforms, it does provide a good insight into the features that MAS developers could reasonably expect from any performance measurement application. The principal characteristics of this structure are as follows:
\begin{itemize}
	\item Primary data channels consist of raw polling sensors at the heart of the system execution engine that gather simple low-impact data elements such as counters and event sensors.
	\item Secondary channels provide more elaborate information, such as summaries of the state of individual components and history analysis that stores performance data over lengthy running times.
	\item Computer-level metrics provide data on such items as CPU load, network load and memory usage.
	\item The message transport service gathers data on messages flowing through it.
	\item An extension mechanism based on servlets allows the addition of visualisation plugins that bind to the performance metrics data source.
	\item The service that is charged with gathering these metrics is designed so as to have no impact on system performance when not in use.
\end{itemize}

Other analysis tools exist for aiding the development of MASs. However, these tend to be narrower in their focus, concentrating only on specific aspects of debugging MASs and typically being only applicable to a specific agent platform. The Agent Factory Debugger \cite{Collier2007} is an example of a tool that is typical of most multi agent frameworks. Its principal function is inspecting the status and mental state of an individual agent: its goals, beliefs, commitments and the messages it has exchanged with other agents. Tools such as this give limited information about the interaction between agents and the consequences of these interactions.

The Brahms toolkit features an AgentViewer that allows developers to view along a time line the actions that particular agents have taken, so as to enable them to verify that the conceptual model of the MAS is reflected in reality~\cite{Seah2005}. An administrator tool for the LS/TS agent platform provides some high-level system monitoring information, such as overall memory consumption and data on the size of the agent population~\cite{Rimassa2005}. Another type of agent debugging tool is the ACLAnalyzer that has been developed for the JADE platform \cite{Botia2004}. Rather than concentrating on individual agents, it is intended to analyse agent interaction in order to see how the community of agents interacts and is organised. In addition to visualising the number and size of messages sent between specific agents, it also employs clustering in order to identify cliques in the agent community.

These latter tools are focused mostly on identifying what actions an agent is carrying out, together with identifying the reasons why such actions are taken (in response to the agents' own belief set or as a result of receiving communication from other agents).

\section{AgentSpotter Overview} \label{sect:overview}

The overriding objective of AgentSpotter is to map the traditional concepts of profiling to agent-oriented concepts so as to build a profiler tool for MAS developers. It could be argued that most mainstream agent toolkits are written in Java, hence the existing profiling tools for the Java programming language are appropriate for the analysis of such platforms and their agents. However, to do so would necessitate the mapping of low-level method profiles to high-level agent-specific behaviour. Thus, tools aimed specifically at Java operate at an inappropriate conceptual level to be of use in agent analysis. Although it may be useful to incorporate some Object-Oriented metrics into an MAS profiling tool, the focus of this paper is on the agent-specific metrics that are appropriate.

Ideally, MASs should be capable of managing their own performance and identifying their own bottlenecks which hamper system efficiency, and indeed much work is being undertaken towards this goal \cite{Horn2001}. However, until this aim is realised, the provision of analysis tools aimed at aiding human developers identify issues with their systems remains of paramount importance.

This section outlines the abstract infrastructure of the AgentSpotter system, which is capable of being integrated into any agent platform. An analysis of the integration of AgentSpotter into a specific agent platform (namely Agent Factory) is contained in Section~\ref{sect:integration}.

\begin{figure}[!t] \caption{AgentSpotter abstract architecture}
\label{fig:as_abstract}
\begin{center}
\includegraphics[scale=0.55]{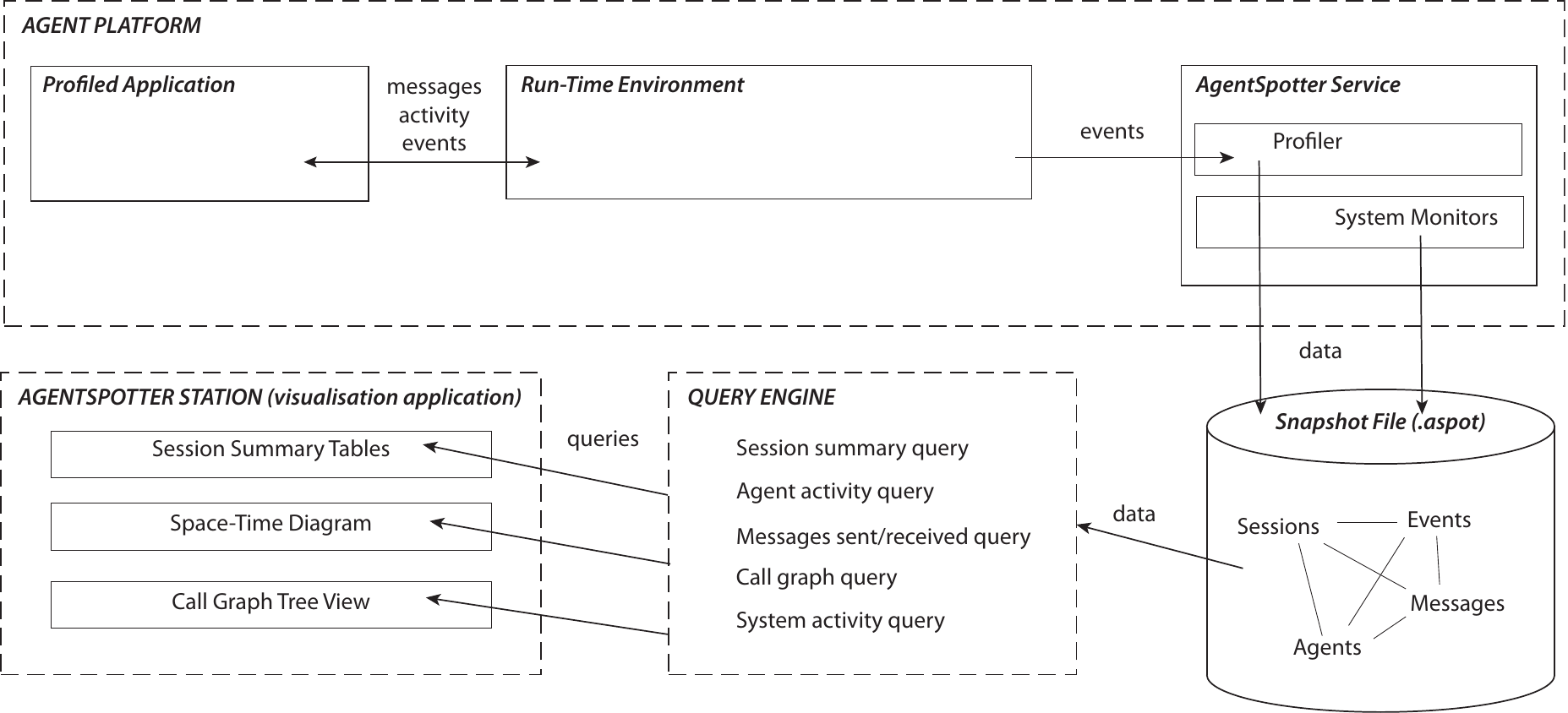}
\end{center}
\end{figure}

The AgentSpotter abstract architecture is displayed in Figure~\ref{fig:as_abstract}, using the following graphical conventions:
\begin{itemize}
\item Top-level architectural units are enclosed in dashed lines and are titled in slanted capital letters, e.g. \dashbox{4}(160, 18){\textbf{\emph{AGENT PLATFORM}}}%\end{picture}
\item Self-contained software packages are enclosed in solid lines e.g \fbox{Profiler}
\item Logical software modules (groups of packages) are titled using slanted capitalised names e.g. \fbox{\textbf{\emph{AgentSpotter Service}}}
\item Arrows denote data or processing interactions e.g.
\begin{picture}(60,15)
\put(10,0){queries}
\put(0,-5){\vector(1,0){60}}
\end{picture}
\end{itemize} 

At the highest level, the \emph{AgentSpotter Service} should communicate with the \emph{Run-Time Environment} to capture the profiling data from a \emph{Profiled Application} running inside an \emph{Agent Platform}. The captured data should be stored into a \emph{Snapshot File} which would then be processed by a \emph{Query Engine} to generate the input data for \emph{AgentSpotter Station}, the visualisation application.

The AgentSpotter profiler monitors performance events generated by the Agent Platform's Run-Time Environment. These include such events as agent management events, messaging and other platform service activity. Additionally, the AgentSpotter service may employ system monitors to record performance information such as CPU load, memory usage or network throughput. This provides a general context for the event-based information.

The event data and other information collected by the AgentSpotter profile is stored in a \emph{snapshot file}, which contains the results of a single uninterrupted data capture session. This snapshot contains a series of raw instrumentation data. Because large MASs may generate potentially hundreds of events per second, it is necessary to introduce a \emph{Query Engine} that is capable of extracting summaries and other information and make it available to visualisation tools in a transparent manner. Ideally, this should be through a data manipulation language such as SQL so as to facilitate the specification of rich and complex queries.

The final component of the abstract architecture is the \emph{AgentSpotter Station}, which is the visualisation tool that summarises the information gathered from the Query Engine in a visual form. The principal focus of this paper is the Space-Time Diagram, which is presented in Section~\ref{sect:spacetime}.

When profiling any application, it is important to identify the appropriate execution unit for profiling (e.g. in the context of object-oriented programming, this would typically be an object or a method). For profiling a MAS, we believe that the appropriate execution units are individual agents.

At the \emph{agent level}, when considering only the autonomous computational entity abstracted away from the interaction with its peers, the focus is on responsiveness. It is obvious that the main influence on the responsiveness of an agent is the amount of computation it requires to carry out the set of tasks required to meet its design objectives. Agreeing with the BT researchers in~\cite{Lee98scal}, we make a further distinction based on the \emph{rationality level} of the agent architecture. For reactive agents, this set of tasks includes only purely reactive behaviour, which is similar to the behaviour of a traditional process. For deliberative agents that implement a reasoning system, the computational cost of the reasoning activity must be considered separately from the actual task execution. Based on this analysis, the minimum information provided by the system should include:

\begin{itemize}
\item \textbf{Agent description:} name, role, type (deliberative or reactive) of the agent.
\item \textbf{Cumulative activity:} cumulative computation time used by the agent.
\item \textbf{Perception time:} perception time used by a deliberative agent.
\item \textbf{Action time:} task execution time used by a deliberative agent.
\item \textbf{Reasoning time:} reasoning time used by a deliberative agent.
\item \textbf{\% session activity:} percentage of the global session computation time used by the agent.
\item \textbf{Number of iterations:} number of non-zero duration iterations used by the agent.
\item \textbf{Total number of messages sent and received:} total number of ACL messages exchanged by the agent.
\end{itemize}

Additionally, a number of global statistics should also be maintained:
\begin{itemize}
\item \textbf{Total duration:} session run-time recorded.
\item \textbf{Total activity:} amount of computation time recorded over the session.
\item \textbf{Total number of messages:} number of messages sent or received by agents on the platform being profiled.
\item \textbf{Average number of active agents per second:} This gives an idea of the level of concurrency in the application.
\end{itemize}

\begin{table}[p]
    \caption{Benchmark application flat profile}

\begin{center}

\begin{tabular}{ p{6cm} p{2cm} }

Total Session Time  & 18:50.691 \\
Total Activity      & 10:29.164 \\
Messages Sent       & 1206 \\
Messages Received   & 1206 \\
Time Slice Duration & 1000 ms \\

\\
\\
\end{tabular}

\scalebox{0.95}{%
  \begin{tabular*}{18cm}{ p{1.8cm} *{8}{r} }

Agent & $T>0$                 & $T>100\%$           & Activity & \% Session          & $Max(T)$ & $Average(T)$ & Msg.   & Msg. \\
      & {\small iterations }  & {\small overload } & mm:ss.ms &   {\small activity}  & ss.ms       & ss.ms            & sent & rec. \\

agent001 & 338 & \fbox{22} & \fbox{1:08.564} & \fbox{10.90} & \fbox{3.740} & \fbox{0.202} & 6 & 57 \\
agent009 & 365 & 21 & 1:04.257 & 10.21 & 3.425 & 0.176 & 13 & 77 \\
agent004 & 349 & \fbox{22} & 1:01.529 & 9.78 & 3.235 & 0.176 & 10 & 69 \\
agent014 & 284 & 14 & 46.413 & 7.38 & 3.148 & 0.163 & 2 & 36 \\
agent003 & 401 & 13 & 43.881 & 6.97 & 3.323 & 0.109 & 12 & 76 \\
agent006 & 361 & 12 & 40.141 & 6.38 & 3.279 & 0.111 & 12 & 73 \\
agent005 & 367 & 12 & 34.903 & 5.55 & 3.325 & 0.095 & 17 & 76 \\
agent013 & 301 & 9 & 34.716 & 5.52 & 3.190 & 0.115 & 14 & 71 \\
agent007 & 378 & 11 & 31.864 & 5.06 & 3.356 & 0.084 & 21 & 71 \\
agent008 & 357 & 7 & 30.850 & 4.90 & 3.201 & 0.086 & 14 & 72 \\
agent010 & 330 & 8 & 30.280 & 4.81 & 3.147 & 0.091 & 21 & 81 \\
agent015 & 285 & 9 & 29.382 & 4.67 & 3.257 & 0.103 & 4 & 42 \\
agent002 & 348 & 8 & 23.196 & 3.69 & 3.147 & 0.066 & 9 & 70 \\
agent011 & 357 & 5 & 19.363 & 3.08 & 3.095 & 0.054 & 4 & 39 \\
agent012 & 225 & 3 & 13.172 & 2.09 & 3.049 & 0.058 & 9 & 41 \\
master2 & \fbox{901} & 0 & 6.681 & 1.06 & 0.183 & 0.007 & 504 & \fbox{86} \\
master1 & 873 & 0 & 6.485 & 1.03 & 0.227 & 0.007 & \fbox{514} & 82 \\
agent024 & 46 & 2 & 6.281 & 1.00 & 3.045 & 0.136 & 3 & 7 \\
agent019 & 31 & 1 & 4.449 & 0.71 & 3.014 & 0.143 & 0 & 5 \\
agent026 & 42 & 1 & 4.400 & 0.70 & 3.084 & 0.104 & 0 & 4 \\
agent030 & 26 & 1 & 4.002 & 0.64 & 3.132 & 0.153 & 2 & 8 \\
agent017 & 46 & 1 & 3.811 & 0.61 & 3.031 & 0.082 & 0 & 3 \\
agent025 & 40 & 1 & 3.767 & 0.60 & 3.006 & 0.094 & 0 & 3 \\
agent027 & 31 & 1 & 3.694 & 0.59 & 3.103 & 0.119 & 0 & 2 \\
agent018 & 38 & 1 & 3.384 & 0.54 & 3.044 & 0.089 & 2 & 7 \\
agent020 & 39 & 0 & 1.762 & 0.28 & 0.547 & 0.045 & 0 & 3 \\
agent022 & 47 & 0 & 1.523 & 0.24 & 0.559 & 0.032 & 5 & 13 \\
agent021 & 32 & 0 & 1.300 & 0.21 & 0.555 & 0.040 & 2 & 7 \\
agent016 & 219 & 0 & 1.194 & 0.19 & 0.555 & 0.005 & 2 & 6 \\
agent029 & 38 & 0 & 1.039 & 0.17 & 0.550 & 0.027 & 2 & 8 \\
agent028 & 45 & 0 & 0.749 & 0.12 & 0.546 & 0.016 & 1 & 4 \\
agent032 & 34 & 0 & 0.749 & 0.12 & 0.561 & 0.022 & 1 & 4 \\
agent031 & 36 & 0 & 0.742 & 0.12 & 0.545 & 0.020 & 0 & 2 \\
agent023 & 40 & 0 & 0.598 & 0.10 & 0.543 & 0.014 & 0 & 1 \\
agent033 & 30 & 0 & 0.043 & 0.01 & 0.003 & 0.001 & 0 & 0 \\
\\
\\

  \end{tabular*}}
\end{center}
 \label{tab:benchmark-flat}

\end{table}

Following the convention of traditional profiling tools, we describe this information as a \emph{flat profile}. AgentSpotter displays this by means of a JTable (provided by Java's Swing interface tools). An example of how the information is presented is given in Table~\ref{tab:benchmark-flat}. This is not necessarily an exhaustive list of every piece of information a developer may desire for identifying problems with a system, however we believe that other metrics (such as memory consumption) are not as crucial for the purposes of profiling the application.

\section{Agent Factory Integration} \label{sect:integration}

\begin{figure}[!h] \caption{AgentSpotter concrete architecture}
\label{fig:as_concrete}
\begin{center}
\includegraphics[scale=0.55]{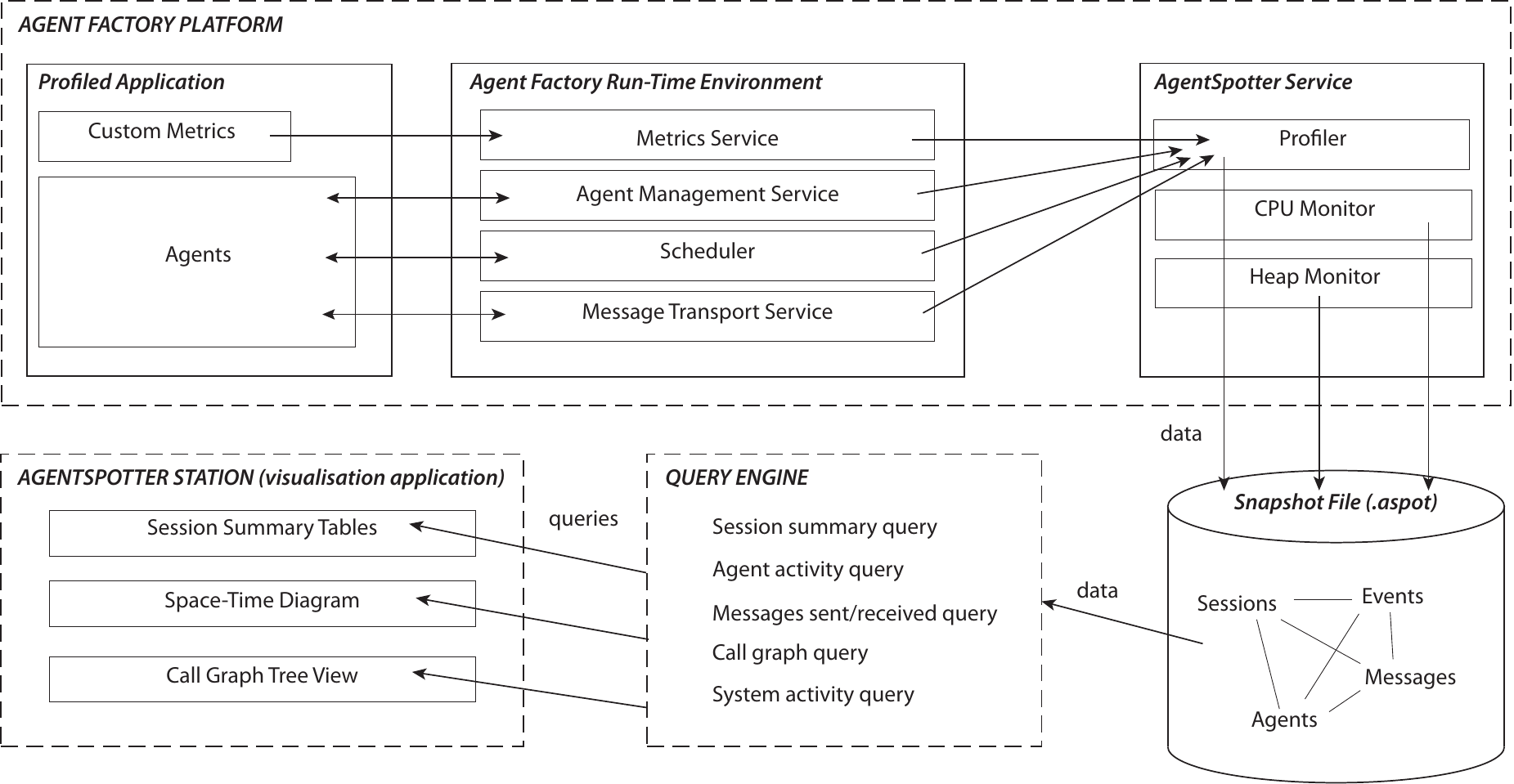}
\end{center}
\end{figure}

Following the definition of the abstract architecture outlined above, a concrete (i.e. platform-specific) implementation was created for Agent Factory. Agent Factory is a cohesive framework that supports a structured approach to the development of agent-oriented applications \cite{Collier2003}. This implementation is illustrated in Figure~\ref{fig:as_concrete}, which uses the same graphical conventions as Figure~\ref{fig:as_abstract}.

To create a concrete implementation, only the platform-specific details must change, as the mechanisms required to monitor events vary from one agent platform to another. In contrast, the AgentSpotter file processing and visualisation components (shown in the lower part of Figure~\ref{fig:as_concrete}) are identical to those in the abstract architecture (Figure~\ref{fig:as_abstract}). Thus, when implementing AgentSpotter for a new type of agent platform, only the AgentSpotter Service that is coupled directly with the platform needs to be reprogrammed. Provided this service creates snapshot files in a consistent way, the Query Engine need not differentiate between agent platforms. AgentSpotter snapshot files are actually transportable single-file databases managed by the public domain \emph{SQLite Database Engine}~\cite{sqlite}. As a result, profiling data is stored as queryable relational database tables.

Within the Agent Factory Run-Time Environment, there are three specific subsystems that generate events of interest in agent profiling, and as such are recorded by the AgentSpotter service. First, the \emph{Agent Management Service} is responsible for creating, destroying, starting, stopping and suspending agents. It generates events corresponding to each of these actions, which are recorded in the snapshot file. The \emph{Scheduler} is charged with scheduling which agents are permitted to execute at particular times and generates events based on this. Finally, the \emph{Message Transport Service} records the sending and receipt of FIPA messages by agents.

\begin{figure}[!t]
	\centering
    \includegraphics*[scale=0.64]{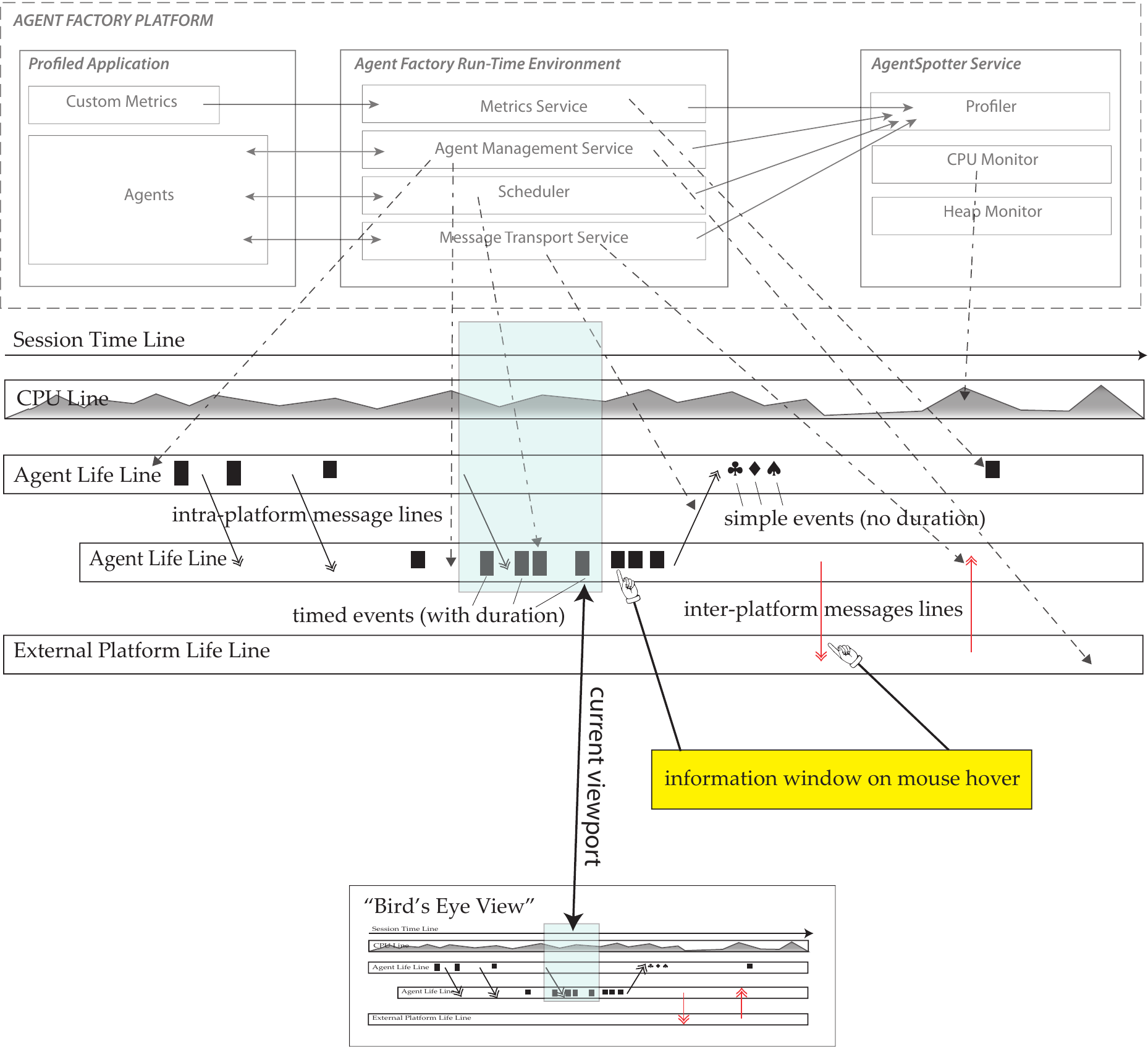}
    \caption[AgentSpotter Space-Time Diagram specification]{AgentSpotter Space-Time Diagram specification annotated with AgentSpotter for Agent Factory infrastructure links (see Figure~\ref{fig:as_concrete}).}
    \label{fig:spacetime}
\end{figure}

\section{Space-Time Diagram} \label{sect:spacetime}

In Section~\ref{sect:overview}, we outlined the minimum amount of information that should be made available by an agent profiler. However, this information can be presented merely by the creation of a simple table. We believe that proper visualisation tools will be far more useful to a developer in understanding a MAS. This section introduces the Space-Time Diagram, which is at the core of the AgentSpotter Station visualisation application. The aim of this diagram is to give as much detail and context as possible about the performance of the MAS to the developer. The user may pan the view around and zoom in and out so as to reveal hidden details or focus on minute details.

The \emph{Session Time Line} represents the running time of the application being profiled. Regardless of the position and scale of the current viewport, this time line remains visible to provide temporal context to the section being viewed and also to allow a developer to move to various points in the session.

The \emph{CPU Line} is a graphical plot of the CPU load of the host system during the session. A vertical gradient going from green (low CPU usage) to red (high CPU usage) provides a quick graphical sense of system load. A popup information window reveals the exact usage statistics once the mouse is hovered over the line.

Perhaps the most important feature of the space-time diagram is the \emph{Agent Time Lines}. Each of these display all the performance and communication events that occur for a single agent during a profiling session. A number of visual features are available to the developer so as to gain greater understanding of the status and performance of the system. For instance, an agent time line begins only at the point in time when the agent is created. This facilitates the developer in viewing the fluctuations in the agent population. Another simple visual aid is that a time line's caption (i.e. the name of the associated agent) is always visible, regardless of what position along the line a developer has scrolled to. Visual clutter may also be reduced by temporarily hiding certain time lines that are not of interest at a particular point in time.

The time line also changes colour to distinguish busier agents from the rest of the community. Darker lines indicate agents that have consumed a greater proportion of the system's CPU time. In a situation where system performance has been poor, this will allow a developer to quickly identify candidate agents for debugging, if they are consuming more resources than is appropriate or expected.

The default ordering of the time lines shares this aim. The time lines are in descending order of total computation time, again visually notifying the developer of those agents consuming more processing resources. However, a developer may alter this default order by dragging time lines into different positions, perhaps to group lines with particularly interesting interactions.

In addition to this simple information, the main purpose of the time line is to show events performed by an agent that are likely to be of interest from a performance point of view. These \emph{performance events} are divided into two categories. \emph{Simple performance events} are those that have a time stamp only. These are shown by means of standard icons (such as an envelope icon to denote that a message was received by the agent). 

\begin{figure}
    \centering
    \includegraphics[scale=0.55]{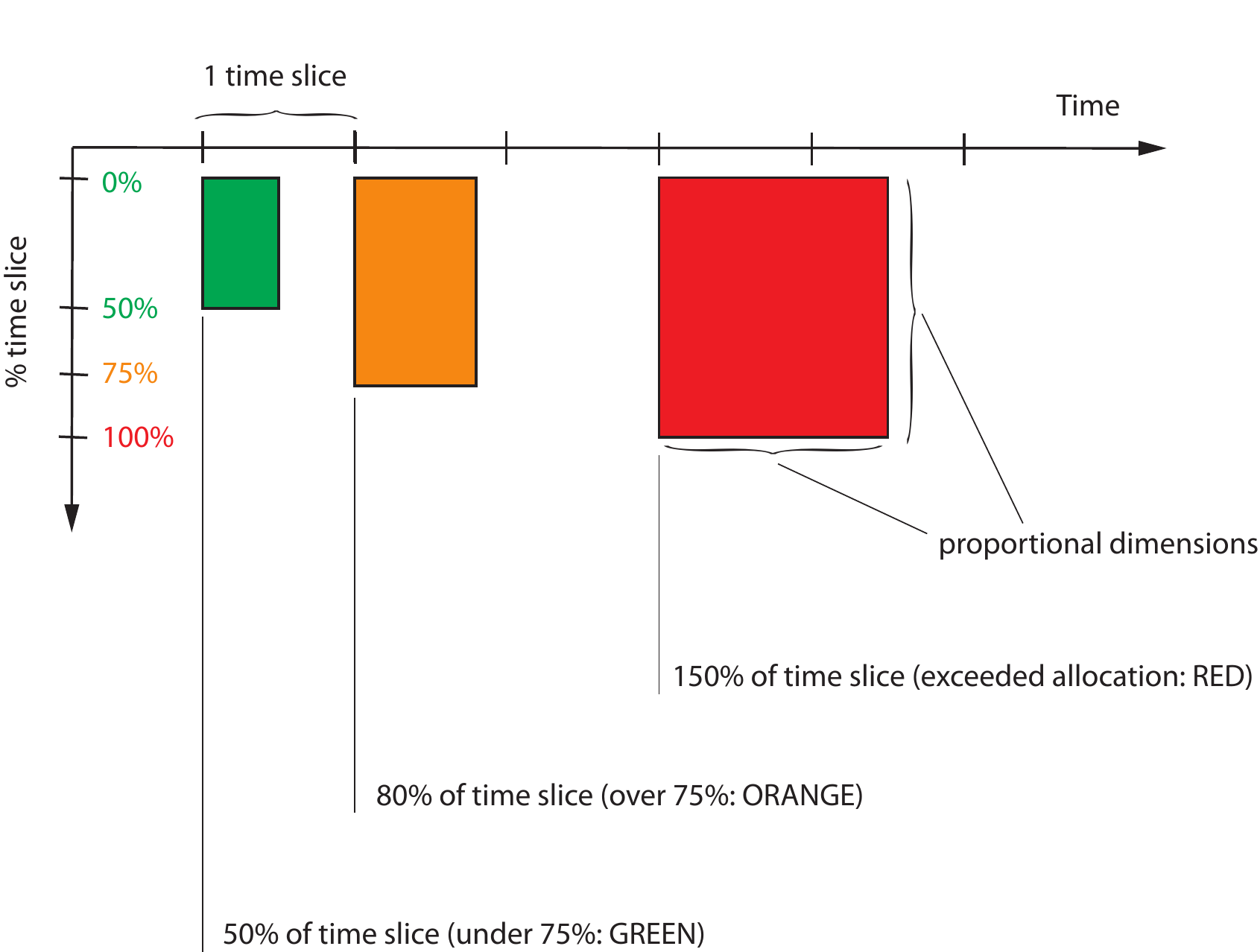}
    \caption{Agent Factory agent activity representation in the Space-Time Diagram}
    \label{fig:slice}  
\end{figure}

The other category of performance events are \emph{timed performance events}. These events are typically actions being performed by an agent. A basic timed performance event displays as a rectangle with a fixed height and a width proportional to its duration. More elaborate event representations can be implemented. For example, in the context of Agent Factory's time slice based scheduler, we have represented the concept of time slice overshoots i.e. when an agent has overused its time slice allocation. 

An example of how these timed events are represented is given in Figure~\ref{fig:slice}. Each agent has a particular time slice within which it is expected to perform all of its actions in a particular iteration. Agents exceeding their time slice may prevent or delay other agents from accessing CPU. As a visual aid to identifying when this situation occurs, timing events are represented by coloured rectangles. The size of these rectangles is proportional to the duration of the event and the percentage of the allocated timeslice used. The colour code also indicates how the agent has used its available time. A green rectangle indicates that the agent has used anything up to 75\% of the time available. From 75\% to 100\%, an orange rectangle indicates that the agent  may require further analysis from the developer to avoid the danger of exceeding the time slice. Finally, whenever an agent exceeds its time, a red rectangle is used. It must be acknowledged that as a result of different approaches to scheduling agent platforms, timed performance events may not be available to a specific AgentSpotter implementation. When this arises, all events will be recorded as simple performance events.

Communication between agents is shown by lines linking the appropriate agent timelines. These include arrows to indicate the direction of the communication. Hovering the mouse pointer over such a line causes a popup window to display the FIPA headers and content of the message. There is also a distinction made between messages passed between agents housed on the same agent platform (intra-platform) and those passed between agents on different platforms (inter-platform). Since agents on other platforms will not have an agent time line, an \emph{external platform life line} is drawn for each other platform with which agents communicate. Rather than linking with individual agent time lines, communications with these platforms are drawn directly to the external platform life line.

The combination of these communication lines and the performance event indicators are very useful in identifying the causes of agent activity. Given the inherently social nature of MASs, it is very common for agent activity to be motivated by communication. For example, an agent may be requested to perform a task by some other agent. Alternatively, an agent may receive a piece of information from another agent that it requires in order to perform a task to which it has previously committed as a result of its own goals and plans.

Providing such detailed visualisation of a MAS requires a substantial amount of screen space. The basic features of zooming and panning are complemented by the provision of a ``bird's eye view'', which displays a zoomed-out overview of the entire session. This allows the user to quickly move the current viewport to focus on a particular point in time during the session, as illustrated in Figure~\ref{fig:spacetime}.

\section{Evaluation} \label{sect:evaluation}

Having outlined the required features of AgentSpotter, along with details of its implementation, it is necessary to demonstrate how it can be utilised on a running MAS. To this end, a specialist benchmark application was developed that will allow the features of the AgentSpotter application to be shown.

\subsection{Specification}
The aim of the benchmark application is to perform all the activities necessary for AgentSpotter to display its features. The requirements for the application can be summarised as follows:
\begin{itemize}
\item \textbf{Load history:} a normally distributed random load history should be generated so that we can get an idea of a ``normal'' profile which can be contrasted with ``abnormal'' profiles where, for example, a single agent is monopolising all the load, or the load is spread equally among all agents.
\item \textbf{Agent population:} the number of active agents should be changeable dynamically to simulate process escalation.
\item \textbf{Interactions:} in addition to direct interactions, the application should exercise some task delegation scenarios. The idea is to generate multiple hops messaging scenarios and see their impact on performance.
\item \textbf{Messages:} agents should generate a steady flow of messages with occasional bursts of intense communication.
\item \textbf{Performance events:} all three performance behaviours described in Section~\ref{sect:overview} should be represented, i.e. green ($t\leq50\%$ time slice), orange ($50\% \leq t \leq 75\%$ time slice), and red ($t > 100\%$).
\end{itemize}

These requirements were satisfied by creating a MAS with overseer agents that request worker agents to execute small, medium or large tasks. Worker agents that have been recently overloaded will simply refuse to carry out the tasks (in a real application they would inform requester about their refusal). From time to time, overseer agents would request agents to delegate some tasks. In this case, worker agents will behave as overseers just for one iteration. A simple interface allows the user to start and pause the process, along with the ability to set the number of active worker agents.

\subsection{Evaluation Scenario and Objective}

The following simple scenario was played out in order to generate a flat profile and space-time diagram.

\begin{enumerate}
\item Start the session with 12 worker agents and 2 overseer agents.
\item After 10 minutes add 15 worker agents to spread the load.
\item After 4 further minutes, suspend the process for 20 seconds.
\item At this point, reduce the number of worker agents to 12.
\item Run for 5 minutes more and then stop the session.
\end{enumerate}

\subsection{Flat profile}

The resulting flat profile of this test is reproduced in Table~\ref{tab:benchmark-flat}. For the reader's convenience, the maximum value for each column is identified by an enclosing box. Overseer agents are called ``master1'' and ``master2''. The worker agents are called ``agent'' followed by a number e.g. ``agent007''.

Firstly, the benchmark appears to make a good job of producing a load history following a normal distribution.

Secondly, we can draw the following conclusions from a quick study of Table~\ref{tab:benchmark-flat}:  
\begin{itemize}
\item The most active agents in terms of number of iterations are the overseer agents, ``master1'' and ``master2'', however in terms of CPU load and overload, three worker agents are topping the list with 30\% of the total activity: ``agent001'', ``agent009'', and ``agent003''.
\item The agents with the highest CPU load also display a high number of time slice overshoots, and a high average time slice duration.
\item As expected, the overseer agents were very busy exchanging messages with the workers. However, it seems that messaging is not CPU intensive. This possibly results from the way in which message sending is implemented, with the CPU load indicated here corresponding to the scheduling of a message for sending, rather than the actual sending of the message. It may be necessary to attach a specialist monitor to the Message Transport Service to gain full information about the impact of sending messages. This causes the activity percentage of the overseer agents to be very low, at only 1\%.
\end{itemize}

\begin{figure}[p]
\begin{center}
    \includegraphics[scale=0.3]{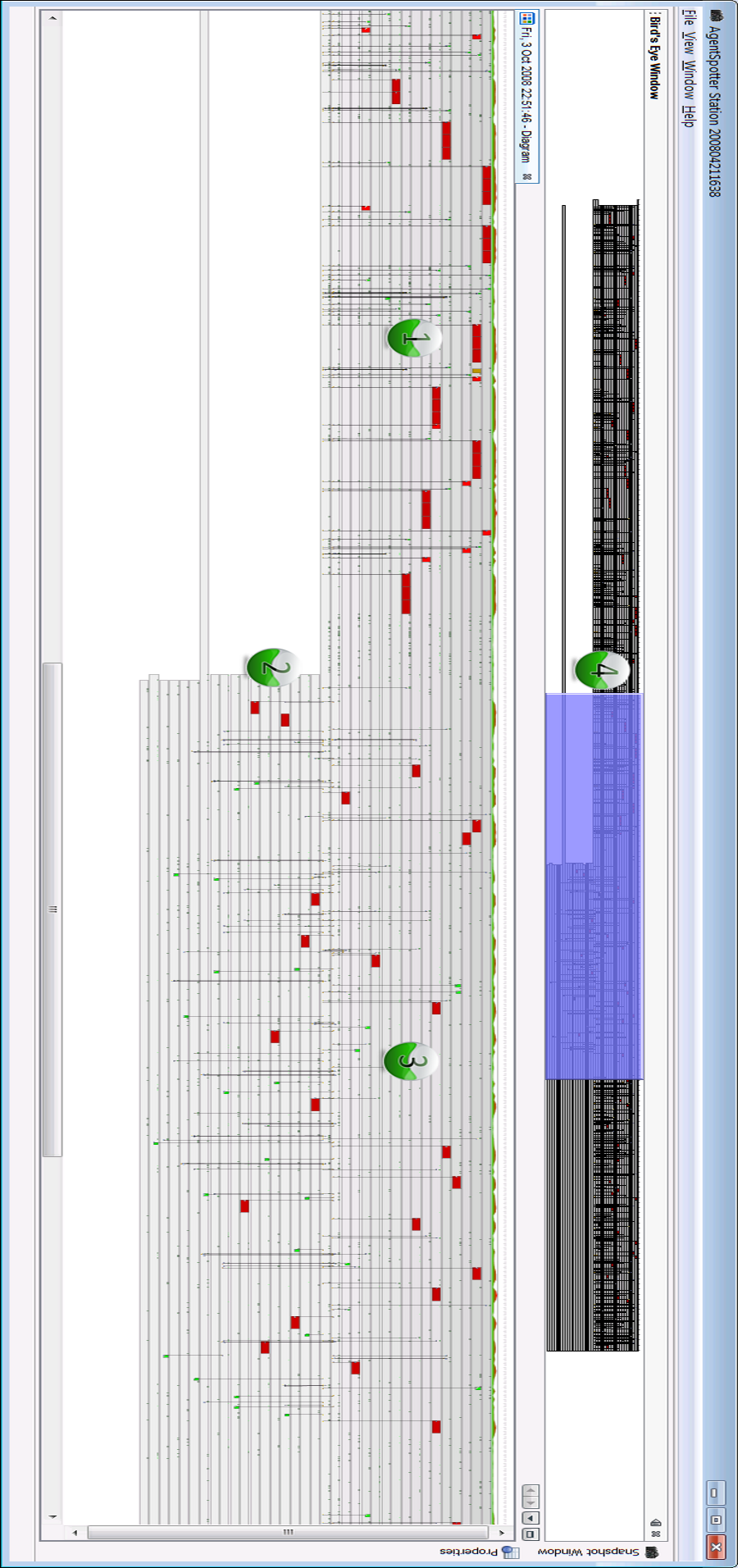}
    \caption{Benchmark application sample space-time diagram (18 minute long session)}
    \label{fig:benchmark-spacetime}
\end{center}
\end{figure}

In this instance, the flat profile lends evidence to the notion that the actual behaviour of the system matches the design principles on which it was built.

\subsection{Space-Time Diagram}

The space-time diagram for this session is shown in Figure~\ref{fig:benchmark-spacetime}. The individual agent time lines can clearly be seen as horizontal bars in the main window of the application. Within these, rectangular boxes represent processing tasks being carried out by each agent. The vertical lines between the time lines represent messages being passed between agents. For this simple scenario, only a single agent platform was used, meaning that there are no external platform time lines to indicate messages travelling to and from other agent platforms. A number of points of interest are labelled on the diagram. These can be described as follows:

\begin{itemize}
\item This portion of the diagram shows what happens when the initial 12 workers are active. The large red rectangles illustrate the time-consuming tasks ordered by the overseers. As mentioned in Section~\ref{sect:spacetime}, these are also identifiable by their size, which increases proportionally to the processing time taken. These blocks never overlap because of the way Agent Factory schedules agents (i.e. agents are given access to the CPU sequentially). It is also noteworthy that the Agent Factory scheduler does not preempt agents that have exceeded their time allocation. The red rectangles also come in bursts, because both overseers send the same order to the same worker at the same time. This was revealed by zooming into what initially appeared to be a single message line. At a high magnification level, there were in fact two messages lines within a few microseconds interval to the same worker.

\item At this point, 15 more workers are added to the system, following a slight pause that is indicated by the temporary absence of message lines. The agent time lines for the additional agents only begin at this point, clearly indicating an increase in the agent population.

\item This third portion shows the impact of the new workers. The red blocks are still present, but they are better spread among the agents, with the new agents taking some of the load from their predecessors.

\item The Bird's Eye View reveals the bigger picture, and reminds us that we are looking only at one third of the overall session.
\end{itemize}

\section{Conclusions and Future Work} \label{sect:conclusions}

Currently, the only concrete implementation of AgentSpotter is for the Agent Factory platform. As noted in Section~\ref{sect:overview}, only the data capture apparatus should require a separate implementation for another platform. It is intended to develop such an implementation for other platforms, such as JADE \cite{Bellifemine1999}.

The most obvious source of improvement for the AgentSpotter application is the addition of extra information above that which is already available. For instance, the performance of additional system services should be recorded, and more details should be collected about agents' performance events, such as the distribution of an agent's execution time among its sensors, actuators, reasoning engine and other components. Finally, the AgentSpotter application currently supports only one agent platform at any given time. The capability to visualise multiple platforms concurrently would be desirable.

\bibliography{master}

\begin{thebibliography}{10}

\bibitem{Knuth1971}
Knuth, D.E.:
\newblock An empirical study of {FORTRAN} programs.
\newblock j-SPE \textbf{1}(2) (April\slash June 1971)  105--133

\bibitem{Graham1982}
Graham, S.L., Kessler, P.B., Mckusick, M.K.:
\newblock Gprof: A call graph execution profiler.
\newblock SIGPLAN Not. \textbf{17}(6) (1982)  120--126

\bibitem{Srivastava1994}
Srivastava, A., Eustace, A.:
\newblock Atom: a system for building customized program analysis tools.
\newblock In: PLDI '94: Proceedings of the ACM SIGPLAN 1994 conference on
  Programming language design and implementation, New York, NY, USA, ACM (1994)
   196--205

\bibitem{jvmti2004}
{Sun Microsystems, Inc.}:
\newblock {JVM Tool Interface (JVMTI), Version 1.0}.
\newblock Web pages at \url{http://java.sun.com/j2se/1.5.0/docs/guide/jvmti/}
  (accessed August 4th, 2008) (2004)

\bibitem{Helsinger2004}
Helsinger, A., Thome, M., Wright, T., Technol, B., Cambridge, M.:
\newblock {Cougaar: a scalable, distributed multi-agent architecture}.
\newblock In: Systems, Man and Cybernetics, 2004 IEEE International Conference
  on. Volume~2. (2004)

\bibitem{Collier2007}
Collier, R.:
\newblock {Debugging Agents in Agent Factory}.
\newblock Lecture Notes in Computer Science \textbf{4411} (2007)  229

\bibitem{Seah2005}
Seah, C., Sierhuis, M., Clancey, W., Cognition, M.:
\newblock {Multi-agent modeling and simulation approach for design and analysis
  of MER mission operations}.
\newblock In: Proceedings of 2005 International conference on human-computer
  interface advances for modeling and simulation (SIMCHI'05). (2005)  73--78

\bibitem{Rimassa2005}
Rimassa, G., Calisti, M., Kernland, M.E.:
\newblock {Living Systems\textregistered Technology Suite}.
\newblock Whitestein Series in Software Agent Technologies and Autonomic
  Computing. In: Software Agent-Based Applications, Platforms and Development
  Kits. {Birkh\"{a}user Basel} (2005)  73--93

\bibitem{Botia2004}
Botia, J., Hernansaez, J., Skarmeta, F.:
\newblock {Towards an Approach for Debugging MAS Through the Analysis of ACL
  Messages}.
\newblock In: Multiagent System Technologies: Second German Conference, MATES
  2004, Erfurt, Germany, September 29-30, 2004: Proceedings, Springer (2004)

\bibitem{Horn2001}
Horn, P.:
\newblock {Autonomic Computing: IBM{'}s Perspective on the State of Information
  Technology}.
\newblock {IBM TJ Watson Labs, NY, 15th October} ({2001})

\bibitem{Lee98scal}
Lee, L.C., Nwana, H.S., Ndumu, D.T., Wilde, P.D.:
\newblock The stability, scalability and performance of multi-agent systems.
\newblock BT Technology Journal \textbf{16}(3) (1998)  94--103

\bibitem{Collier2003}
Collier, R., O{'}Hare, G., Lowen, T., Rooney, C.:
\newblock {Beyond Prototyping in the Factory of Agents}.
\newblock {Multi-Agent Systems and Application III: 3rd International Central
  and Eastern European Conference on Multi-Agent Systems, Ceemas 2003, Prague,
  Czech Republic, June 16-18, 2003: Proceedings} ({2003})

\bibitem{sqlite}
Hwaci:
\newblock Web site for the {SQLite Database Engine} (2008)
  \url{http://www.sqlite.org/} (accessed October, 2008).

\bibitem{Bellifemine1999}
Bellifemine, F., Poggi, A., Rimassa, G.:
\newblock {JADE--A FIPA-compliant agent framework}.
\newblock In: Proceedings of PAAM. Volume~99. (1999)  97--108

\end{thebibliography}

\end{document}